# Preparation of alginate hydrogel microparticles using droplet-based microfluidics: a review of methods

Cheng Zhang, Romain Grossier, Nadine Candoni, Stéphane Veesler

*CNRS, Aix-Marseille Université, CINaM (Centre Interdisciplinaire de Nanosciences de Marseille), Campus de Luminy, Case 913, F-13288 Marseille Cedex 09, France*

## Abstract
This review focuses on a microparticle-producing technique widely used for its efficacy in controlling physicochemical properties: droplet-based microfluidics.
The review describes the various strategies applied under this technique, and the size, shape, concentration, stability and mechanical properties of the alginate hydrogel microparticles obtained.

## Introduction

Hydrogel microparticles are widely used today, especially in biological and pharmaceutical applications. They are usually used as a matrix to encapsulate bioactive agents such as drugs, proteins, cells, etc. (Ahmed, 2015; Maitra and Shukla, 2014) for drug delivery (Agüero et al., 2017), cell culture and tissue engineering. (Utech et al., 2015) Another important use is as cell-mimicking microparticles with similar size, shape, deformability and mechanical properties. (Haghgooie et al., 2010; Merkel et al., 2011; Zhang et al., 2020) Hydrogels can be made of various biopolymers such as gelatin, agarose, alginate, pectin, etc. Alginate stands out because of its low cost, non-toxicity and ease of crosslinking (Lee and Mooney, 2012).

With the increasing interest in alginate hydrogel microparticles, various preparation methods have been reported in the literature, including conventional emulsification (Chan et al., 2002), spray-drying (Santa-Maria et al., 2012), extrusion dripping (Lee et al., 2013), microfluidics (Rondeau and Cooper-White, 2008; Zhang et al., 2006) and soft lithography. (Qiu et al., 2007) The huge diversity of techniques and strategies can make it confusing to choose the right method. The present review focuses on a microparticle-producing technique widely used for its efficacy in controlling physicochemical properties: droplet-based microfluidics. The various strategies applied within this technique and the properties of the microparticles obtained are described in this review.

## I. Alginate hydrogel microparticles

Hydrogels are described as hydrophilic polymeric networks which can absorb and retain large amounts of water within the structure. The hydrogel network is formed by polymer crosslinking. When crosslinking is realized by molecular entanglement, ionic, H-bonding or hydrophobic forces, hydrogels are called physical or reversible gels. Otherwise, when covalent forces intervene, they are called chemical or permanent gels. (Caccavo et al., 2018; Hoffman, 2012)

## I.1. Alginate

Alginate is a natural polysaccharide. Although it can also be synthesized by several bacteria, all the commercially available alginate is produced from the extraction of brown algae (Draget, 2009). Alginate is widely used in the biomedical field because it is biocompatible and non-toxic (Lee and Mooney, 2012).

Sodium alginate (Na-alginate) is the most widely used alginate salt. It dissolves in water to a viscous solution. Alginate is a linear copolymer containing β-D-mannuronate (M) and α-L-guluronate (G) residues (Figure 1).

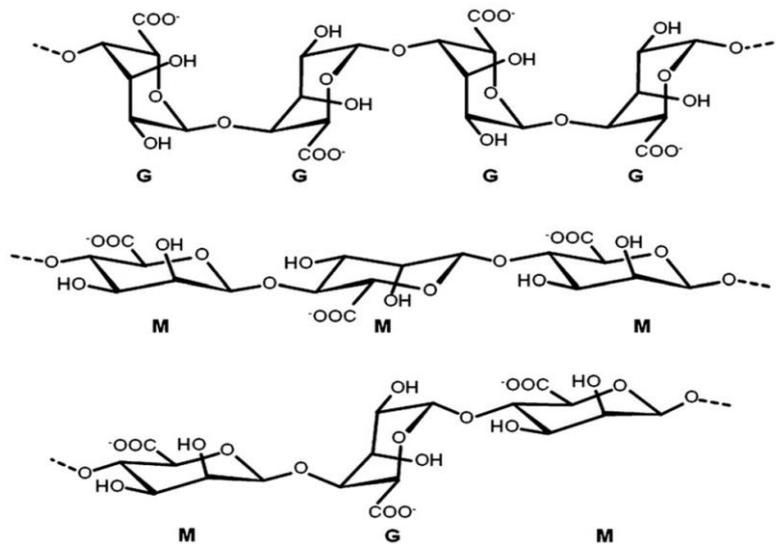

Figure 1. Chemical structures of G-block, M-block and alternating G and M-blocks in alginate. Figure reprinted with permission from Reference (Lee and Mooney, 2012)

## I.2. Gelation of alginate

Alginate hydrogel is produced by gelation which is caused by covalent (Fundueanu et al., 1999) or ionic crosslinking. (Gacesa, 1988; Velings and Mestdagh, 1995) Ionic crosslinking is more commonly used because of its simplicity and mild conditions. It can be carried out at room temperature or up to 100°C, usually with divalent cations as crosslinking agents. Calcium chloride is the most widely used (Lee and Mooney, 2012), due to its non-toxicity (Agüero et al., 2017) and availability.

Only G-blocks (Figure 1) made of consecutive G residues can participate in ionic crosslinking because of their favourable spatial structure (Gacesa, 1988; Lee and Mooney, 2012). Ionic crosslinking of alginate is described by the "egg-box" model (Grant et al., 1973) (Figure 2).

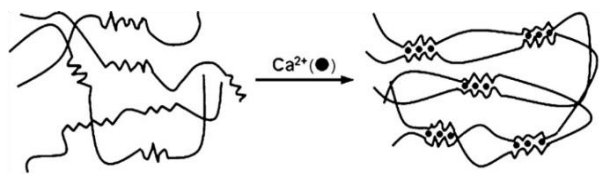

Figure 2. Schematic illustration of the "egg-box" model describing the ionic crosslinking of alginate by calcium cations. Figure reprinted with permission from Reference (Lee and Yuk, 2007)

In this paper, we present a two-step method to produce alginate hydrogel microparticles. First, sodium alginate droplets are generated using droplet-based microfluidics. Second, gelation transforms droplets into alginate hydrogel microparticles via different strategies.

## II. Droplet-based microfluidics

### II.1. Principle of droplet generation

Microfluidics is a technique used to manipulate fluids in channels of micrometric dimensions. Fluids are mixed by adding junctions that connect the channels. When immiscible or partially miscible fluids are mixed in the junction, microdroplets can be generated: this is called droplet-based microfluidics.

The principle is similar to that of conventional emulsification, which consists of blending two immiscible liquids. The advantage of droplet-based microfluidics is monodispersity and repeatability of droplets due to precise control over experimental conditions such as channel geometry, flow rates and viscosities of fluids, etc. (Seemann et al., 2011; Teh et al., 2008) Furthermore, monodisperse droplets can be generated without using surfactant (Liu et al., 2006; Trivedi et al., 2009; Zhang et al., 2020), which is impossible with conventional emulsification.

The droplets generated in droplet-based microfluidics can serve as microreactors to carry out physical, chemical or biological reactions. (Zhu and Wang, 2017) Being small (nL to µL volume), they require a small quantity of reactants. As droplet composition can be made identical, numerous identical experiments can be performed, enabling a reliable statistical approach to data.

### II.2. Flow properties

In droplet-based microfluidics, a continuous fluid and a dispersed fluid are injected separately and then mixed in a junction. Fluids are Newtonian and droplets of the dispersed fluid are generated in the flow of the continuous fluid. The physicochemical properties influencing droplet formation are density, dynamic viscosity, surface tension between the continuous and the dispersed fluids, velocity of the flows and characteristic dimensions of the microfluidic system, such as the diameter of channels (D) for cylindrical microfluidic systems. Based on these properties, fluid dynamics is characterized as follows:

1- Inertial forces and viscous forces are compared through the Reynolds number, calculated using the continuous fluid properties: density ($\rho_C$), dynamic viscosity ($\mu_C$) and flow velocity ($v_C$).

$$Re = \frac{\rho_C \times v_C \times D}{\mu_C} \quad (1)$$

Typically for microfluidics, values of Re are lower than 1: the flow is laminar and the effect of inertia can be ignored. Thus the average velocity v is evaluated from D and the volumetric flow rate Q as follows:

$$v = \frac{Q}{\pi (D/2)^2} \quad (2)$$

2- The generation of droplets in a microfluidic junction creates a free interface between the two fluids, characterized by the interfacial energy $\gamma_{CD}$. The corresponding capillary effects are in competition with gravity effects. The length above which gravity effects dominate capillary effects is the capillary length $l_c$:

$$l_c = \sqrt{\frac{\gamma_{CD}}{\Delta\rho \times g}} \quad (3)$$

with g the gravity acceleration and Δρ the difference in density between the two fluids. For instance, with fluorinated oil FC70 as the continuous fluid and ethanol as the dispersed fluid, $l_c$ is equal to 2.4 mm (Zhang et al., 2015). Hence gravity does not influence the deformation of the interfaces in millimetric or sub-millimetric channels.

3- Shear stress and interfacial energy are compared through the capillary number Ca. When generating droplets of a dispersed fluid in a continuous fluid, Ca is usually calculated using $v_C$ and $\mu_C$ of the continuous fluid, and $\gamma_{CD}$ of the interface between the continuous and the dispersed fluid:

$$\text{Ca} = \frac{\mu_C \times v_C}{\gamma_{CD}} \quad (4)$$

## II.3. Microfluidic geometry

Microfluidic devices can be in the form of either chips with microchannels and junctions produced by soft lithography, or an assembly of capillaries and junctions (Ren et al., 2013). In terms of materials, polydimethylsiloxane (PDMS) is the most commonly used for microfluidic chips (Liu et al., 2006; Zhang et al., 2006). For capillaries, both glass (Baroud et al., 2010; Hu et al., 2012) and fluoropolymer can be used. (Trivedi et al., 2009; Zhang et al., 2020)

The channel geometry of a microfluidic device influences droplet generation. Three frequently used geometries are "cross-flow", "co-flow" and "flow-focusing" (Figure 3).

### II.3.1. Cross-flow

For cross-flow geometry, continuous fluid and dispersed fluid mix with an angle $\vartheta$ (0° < $\vartheta$ ≤ 180°) at the junction (Figure 3 (a)). Where the two fluids meet, first an interface is formed due to the immiscibility of the two fluids. Shear force then pushes the head of the dispersed fluid into the continuous fluid until a part breaks off: the droplet is formed. Then it circulates in the channel of the continuous fluid. (Teh et al., 2008)

Cross-flow geometry is often called T-junction geometry, where two fluids flow orthogonally (Figure 3 (a) i). However, other shapes of junctions can also be used, such as a junction with an arbitrary angle $\vartheta$ (Figure 3 (a) ii), or a Y-shaped junction (Figure 3 (a) iv). For two fluids facing each other ($\vartheta$ = 180°, Figure 3 (a) iii), the geometry is called "head-on". A combination of two junctions (Figure 3 (a) v, vi) can also be used to introduce two different dispersed fluids and one continuous fluid. Cross-flow geometry is widely used due to its ease of assembly and handling. (Seemann et al., 2011; Zhu and Wang, 2017)

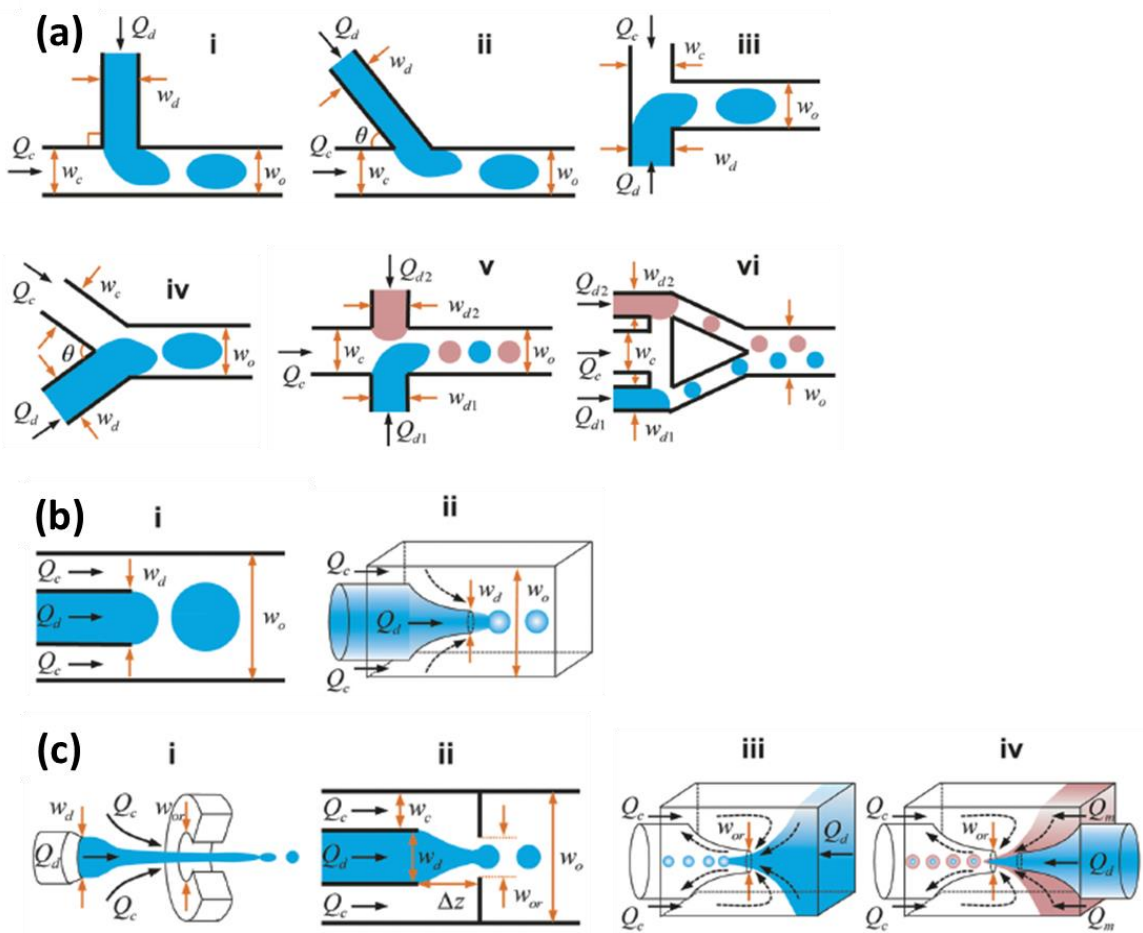

Figure 3. Schematic illustrations (a) "cross-flow", (b) "co-flow" and (c) "flow-focusing" geometries for a microfluidic device. **Q** and **w** denote respectively flow rate and channel width. Subscripts **d**, **c**, **o** and **or** denote respectively dispersed fluid, continuous fluid, outlet channel and orifice. Figure reprinted with permission from Reference (Zhu and Wang, 2017)

### II.3.2. Co-flow
For co-flow geometry, two immiscible fluids flow in two concentric channels (Figure 3 (b)). Droplets are formed at the outlet of the inner channel.

### II.3.3. Flow-focusing
Flow-focusing geometry is actually similar to co-flow geometry. The distinction presented in the literature (Zhu and Wang, 2017) is somewhat ambiguous, leading some to consider flow-focusing as a special co-flow geometry. (Seemann et al., 2011) For flow-focusing geometry, two immiscible fluids are focused through an orifice, which allows smaller droplets to be generated than with co-flow geometry.

### II.4. Droplet generation regime
For each geometry, droplets can be generated following three different break-off mechanisms. The transition from one mechanism to another can be achieved by varying capillary numbers

Ca. (Zhu and Wang, 2017) Figure 4 shows an example of three mechanisms for a cross-flow geometry.

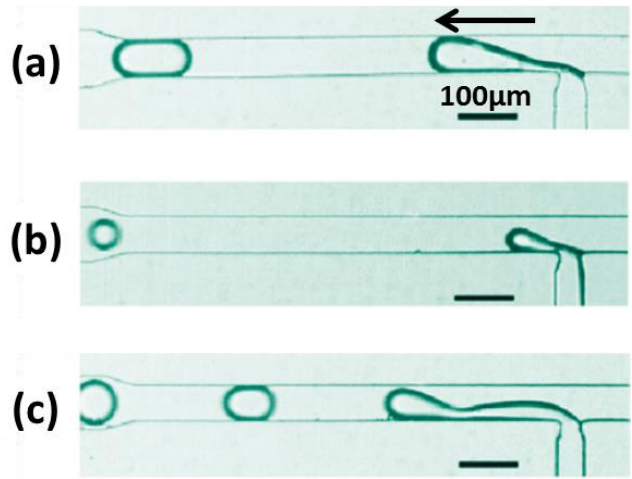

Figure 4. Three break-off mechanisms of droplet generation with a cross-flow geometry: (a) squeezing, (b) dripping and (c) jetting. The arrow indicates the droplet flow direction. Figure reprinted with permission from Reference (Zagnoni et al., 2010). Copyright 2010 American Chemical Society.

### II.4.1. Squeezing
As Figure 4 (a) shows, as it is injected into the principal channel, the dispersed fluid is pushed forward by the continuous fluid. A thin "neck" is thus formed. Because the continuous fluid applies weak shear force, the forming droplet reaches the opposing channel wall without breaking off. The neck becomes thinner until it breaks, so that a plug-shaped droplet confined by channel wall is formed. Squeezing mechanism appears when Ca is low (Ca ≤ 0.01). (De Menech M. et al., 2008)

### II.4.2. Dripping
As Figure 4 (b) shows, the shear force applied is now higher. The forming droplet breaks off before touching the opposing channel wall. A spherical droplet is formed with a diameter smaller than that of the channel. This dripping mechanism appears at a higher Ca (Ca ≥ 0.02). (De Menech M. et al., 2008)

### II.4.3. Jetting
As Figure 4 (c) shows, a liquid jet is emitted from the dispersed fluid channel. It flows and remains attached to the channel wall, due to a strong shear force from the continuous fluid. (Christopher and Anna, 2007) The jet breaks up into droplets at the end because of Rayleigh-Plateau instability. (Zhu and Wang, 2017) Droplets of polydisperse sizes are formed. This jetting mechanism appears at the highest Ca (Ca ≈ 0.2).

## III. Gelation

### III.1. Internal gelation
For internal gelation, crosslinking agents are either soluble or insoluble/slightly soluble in water.

### III.1.1. Water-soluble crosslinking agents

With water-soluble crosslinking agents such as calcium chloride or barium chloride, alginate is crosslinked directly at the interior of droplets. These agents are mixed with Na-alginate before or after droplet generation, using different strategies, as presented below in detail.

#### III.1.1.1. Mixing crosslinking agents before droplet generation

Trivedi et al. worked on cell encapsulation by alginate hydrogel microparticles. (Trivedi et al., 2009) For the preparation of microparticles, an aqueous solution of cell-containing Na-alginate (1 %) and a solution of barium chloride (50 mM) were injected into the capillary and mixed via a Y-shaped junction (Figure 5 (a)). At the exit from the mixing region, highly viscous silicone oil (10 centistokes) without surfactant was injected by flow-focusing in order to generate droplets. However, the mixing of Na-alginate and barium cations triggered ionic crosslinking, causing gelation which impacted droplet generation. Finally, instead of generating droplets as expected, a jet of gel was produced with a partially formed droplet head and a long gelatinous tail (Figure 5 (b)).

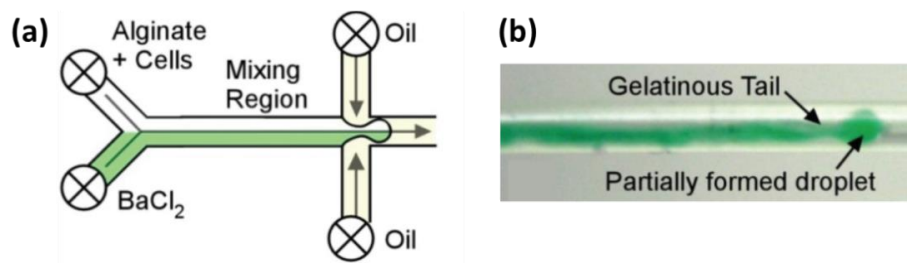

Figure 5. (a) Schematic diagram of mixing Na-alginate/cells and barium chloride solutions before droplet generation in a microfluidic device assembled from fluoropolymer capillaries and junctions. (b) Image of the jet of gel produced in oil. Figure reprinted from Reference (Trivedi et al., 2009)

The problem can be solved by using low concentrations of Na-alginate and calcium chloride solutions. In this case, gelation proceeds after droplet generation and is enhanced by using partially miscible fluids. Rondeau and Cooper-White used dimethyl carbonate (DMC) as the continuous fluid (Rondeau and Cooper-White, 2008) (Figure 6). The solubility of water in DMC is about 3 wt% at room temperature. (Stephenson and Stuart, 1986) Aqueous solutions of Na-alginate (0.5 wt%) and calcium chloride (0.25 wt%) were injected respectively from inlets A and B (Figure 6 (a)). After a short pre-gelation channel, DMC was injected from inlet C. Na-alginate/$CaCl_2$ droplets were generated in DMC (no mention of surfactant usage) by flow-focusing. Along the serpentine channel, because of the low solubility of water in DMC, water diffused gradually from droplets into DMC, causing the shrinkage of droplets along the channel. Internal gelation occurred at the same time. Microparticles with a diameter of 20µm were observed at the outlet of the channel and collected in an aqueous solution of calcium chloride (2N) to reinforce the gelation (Figure 6 (b)). The diameter of Ca-alginate hydrogel microparticles was influenced by the experimental parameters such as the initial concentration of Na-alginate, flow rates of fluids and channel size.

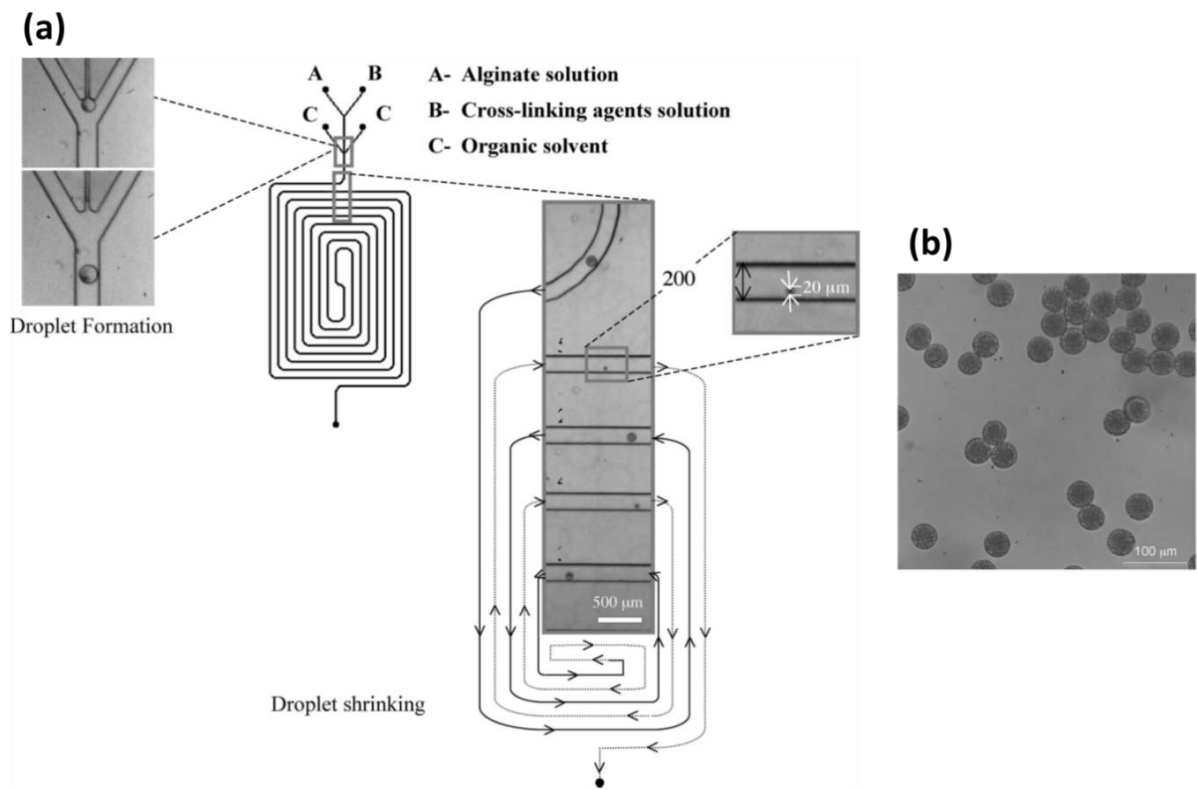

Figure 6. (a) Schematic diagram of a PDMS-based microfluidic device using DMC as the continuous fluid in which water is partially soluble. Droplet shrinkage is observed for an initial concentration of Na-alginate of 0.5 wt%. (b) Micrograph of Ca-alginate hydrogel microparticles collected in an aqueous solution of calcium chloride. Figure reprinted with permission from Reference (Rondeau and Cooper-White, 2008). Copyright 2008 American Chemical Society.

Following the work of Rondeau and Cooper-White, we tested, in a T-junction (Figure 7), the direct generation of droplets of Ca-alginate in dimethyl carbonate (DMC) without surfactant from a mixture of more diluted Na-alginate and calcium chloride solutions (both at 0.06 wt% after mixing). However, this solution was not clear and local gelification was occasionally observed with the naked eye. When these gels entered the channel, droplets were generated in a discontinuous way. This indicates that, even at very low concentrations, thorough mixing of Na-alginate and calcium chloride solutions leads to gelation, disturbing droplet generation.

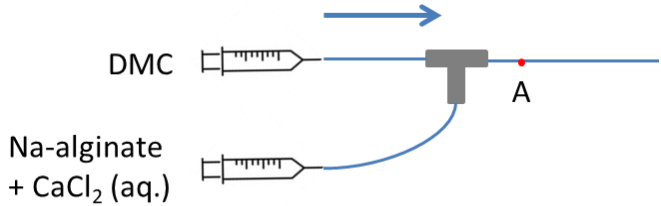

Figure 7. Schematic diagram of the generation of droplets of Ca-alginate with DMC as the continuous fluid and an aqueous mixed solution of Na-alginate and $CaCl_2$ as the dispersed fluid. The arrow indicates the flow direction.

To deal with this issue, Zhang et al. experimented with reducing the mixing region before droplet generation. (Zhang et al., 2006) Using a 5-channel microfluidic device, they mixed Na-alginate fluid (0.5 wt%), calcium chloride fluids (0.1 wt%) and mineral oil fluids with a surfactant (Span 80, no concentration mentioned) as shown in Figure 8. Droplets were generated by co-flow. However, instead of producing discrete droplets, a line of knots connected with each other was formed. This phenomenon persisted with a wide range of flow rates of oil due to viscosity which increased instantly when Na-alginate and calcium chloride were mixed, because of rapid gelation. It was therefore difficult to generate droplets at the junction, despite the use of surfactant.

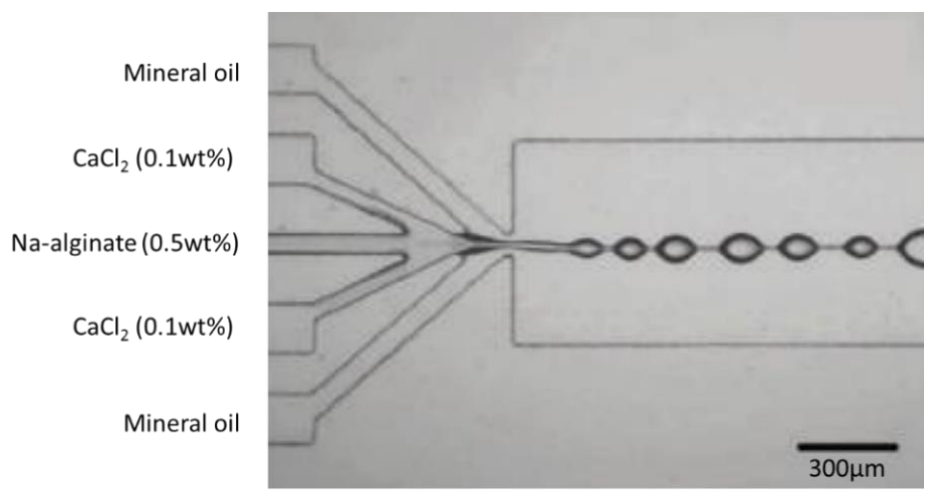

Figure 8. Image of connected knots formed after mixing Na-alginate and calcium chloride solutions in mineral oil with surfactant, in a PDMS-based microfluidic chip. Figure reprinted with permission from Reference (Zhang et al., 2006). Copyright 2006 American Chemical Society.

In a microfluidic device (Figure 9) of similar design to Zhang et al., we were able to generate discrete droplets by using extremely diluted solutions of Na-alginate (0.006 wt%) and calcium chloride (0.002 wt%). The continuous fluid was dimethyl carbonate (DMC) without surfactant. Droplets were observed after the cross-junction (point A in Figure 9 (a)). Since they were relatively close to each other in the channel, causing coalescence at the outlet (point B in Figure 9 (a)), a second flow of DMC was introduced as a spacer using a T-junction. When the second DMC flow rate was relatively low, the generation of droplets upstream was not disturbed, so that droplets were uniform (Figure 9 (b)). However, the coalescence at the outlet persisted. Thus, high second DMC flow rates were applied to sufficiently increase the distance between droplets. Nevertheless, this quickly disturbed the generation of droplets upstream, as indicated by heterogeneities in droplet size and frequency (Figure 9 (c)). Using surfactant would solve this problem.

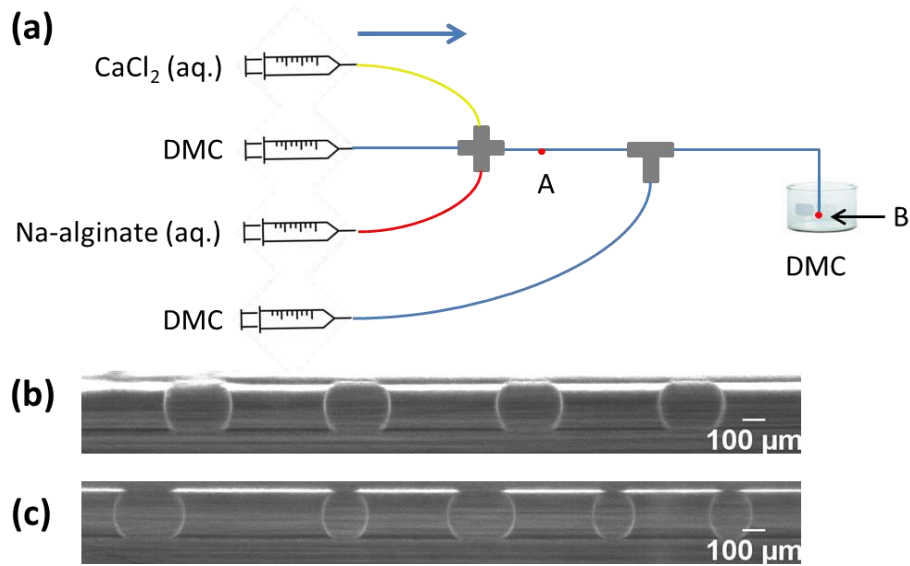

Figure 9. (a) Schematic diagram of generating Ca-alginate droplets in dimethyl carbonate (DMC) using a cross-junction. The T-junction served to introduce DMC as a spacer to increase the distance between droplets. Micrograph of droplets observed at point A when (b) the generation of droplets was not disturbed by introducing the spacer and (c) when it was disturbed.

Thus, in most cases where immiscible fluids are used, mixing Na-alginate and water-soluble crosslinking agents prior to droplet generation is not an efficient approach due to gelation. This problem can be solved by using partially miscible fluids, which makes it possible to apply lower concentrations of Na-alginate and water-soluble crosslinking agents. However mixing should be limited to achieve incomplete pre-gelation. Another possible solution consists of using extremely diluted aqueous solutions of Na-alginate and crosslinking agents, which requires the use of surfactant to avoid coalescence. Other strategies involve performing the gelation after the droplet generation.

### *III.1.1.2. Mixing crosslinking agents after droplet generation*

Xu et al. prevented rapid gelation by delaying the direct contact between Na-alginate and calcium cations. (Xu et al., 2008) In a first cross-junction, two face-to-face channels were used to introduce calcium chloride (2 wt%) and Na-alginate (2 wt%) solutions (Figure 10 (a)) perpendicularly to a flow of water. Thus, after the first cross-junction, a flow of water (acting as a buffer) separates the flows of Na-alginate and calcium chloride. Then octyl alcohol oil (no mention of surfactant) was injected at a second cross-junction. Droplets of Na-alginate/$CaCl_2$ were generated by flow-focusing. In the "synthesizing channel" (Figure 10 (a)), within each droplet, mixing Na-alginate and calcium chloride induced internal gelation. In this way droplets were transformed into Ca-alginate hydrogel microparticles (Figure 10 (b)).

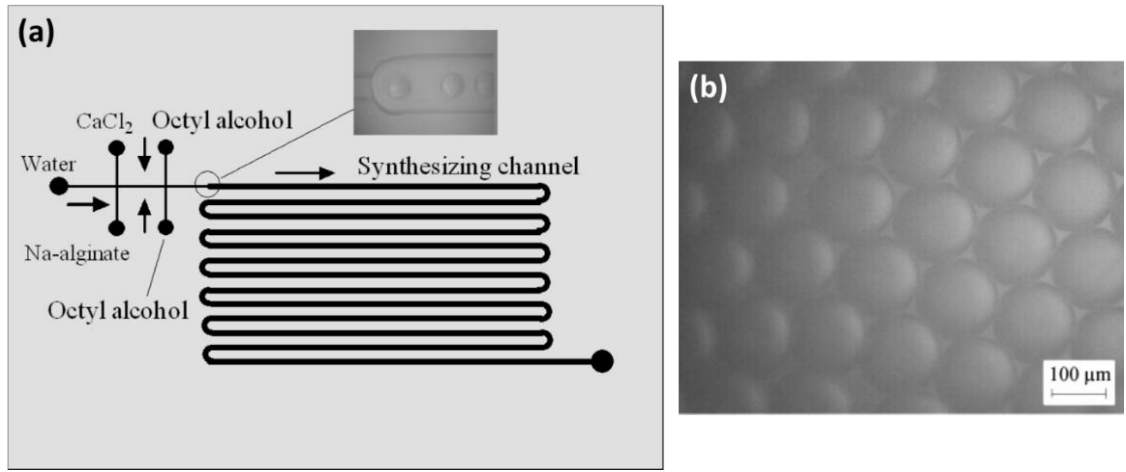

Figure 10. (a) Schematic diagram of Ca-alginate hydrogel microparticles prepared in a poly(methyl methacrylate) (PMMA) based microfluidic device. (b) Micrographs of Ca-alginate hydrogel microparticles. Figure reprinted with permission from Reference (Xu et al., 2008)

Another strategy to delay gelation was carried out by Liu et al. (Liu et al., 2006) involving coalescence of Na-alginate droplets with calcium chloride droplets generated separately. First, on a microfluidic chip (Figure 11 (a)), Na-alginate (2 wt%) droplets (Figure 11 (b)) and calcium chloride (2 wt%) droplets (Figure 11 (c)) were generated in soybean oil without surfactant by flow-focusing using two independent cross-junctions. Then droplets converged via a T-junction (Figure 11 (d)) followed by two successive circular expansion chambers (Figure 11 (d, e)). Thus, droplets could collide either at the T-junction or in circular chambers. Within the coalesced droplets, Na-alginate was crosslinked by calcium cations forming Ca-alginate hydrogel microparticles. With different flow rates and channel geometries, various shapes and sizes of microparticles could be produced (Figure 11 (f)).

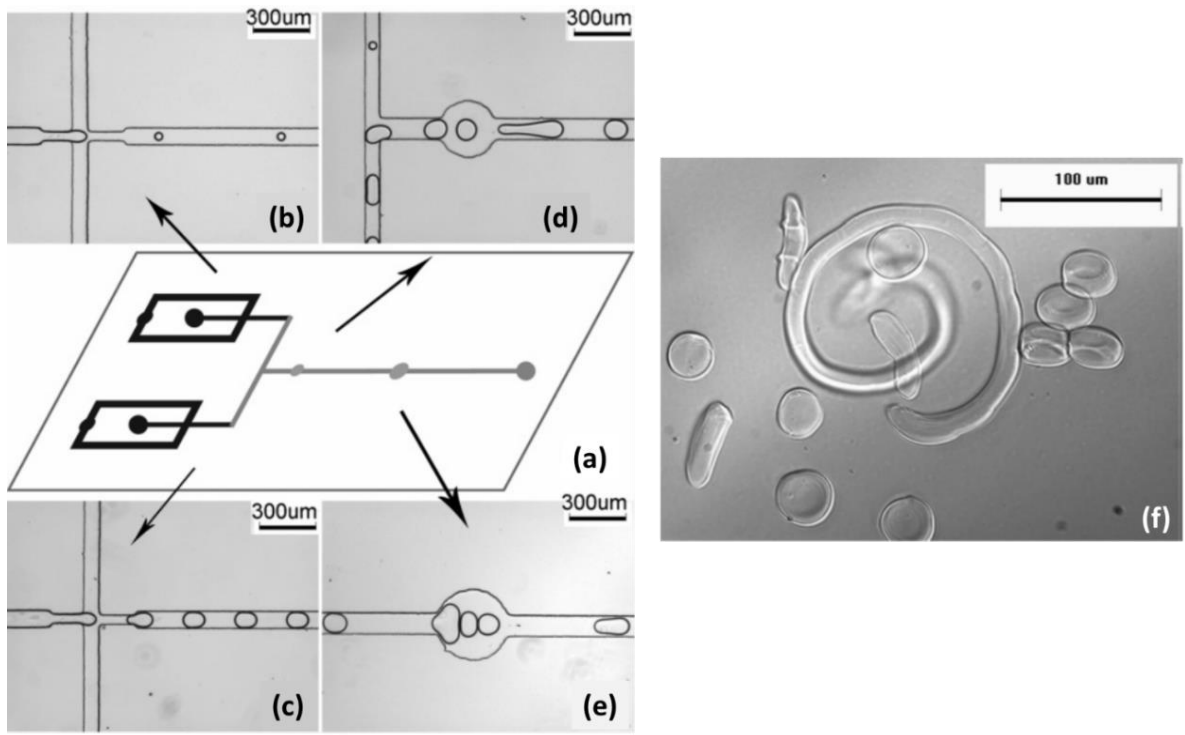

Figure 11. (a) Schematic diagram of the PDMS-based microfluidic device. (b) Flow-focusing channel to generate alginate droplets. (c) Flow-focusing channel to generate $CaCl_2$ droplets. (d) T-junction followed by a first circular expansion chamber. (e) A second circular expansion chamber. (f) Ca-alginate hydrogel microparticles of different shapes and sizes. Figure reprinted with permission from Reference (Liu et al., 2006). Copyright 2006 American Chemical Society.

Droplets could also be coalesced by exploiting physicochemical parameters between the continuous fluid and the dispersed fluid. (Trivedi et al., 2010; Trivedi et al., 2009) As shown in Figure 12 (a), droplets of Na-alginate (1 wt%) containing cells were generated upstream in a highly viscous silicone oil (10 centistoke) by flow-focusing without surfactant. An aqueous solution of barium chloride (50 mM) was injected downstream by a T-junction. With the help of dye, observations at the T-junction indicated that barium chloride fluid merged spontaneously with Na-alginate/cells droplets (Figure 12 (b)), instead of forming independent barium chloride droplets. However, when using low-viscosity and low-interfacial energy $\gamma_{CD}$ soybean oil, independent droplets of barium chloride were observed (Figure 12 (c)). They coalesced downstream with Na-alginate/cells droplets. This implied that successful coalescence of droplets could only take place with appropriate interfacial energy and viscosity (Trivedi et al., 2009)

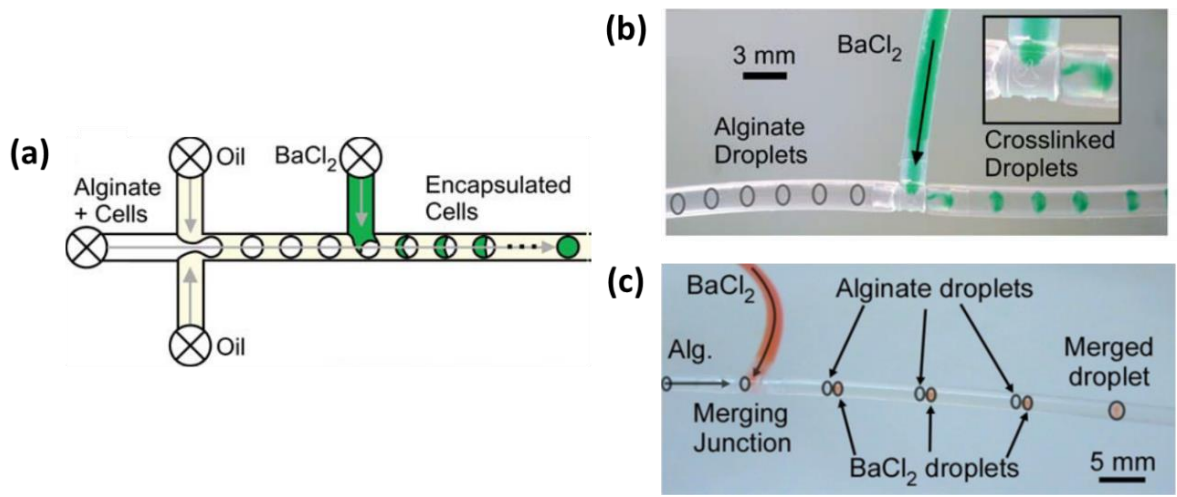

Figure 12. (a) Schematic diagram of Ba-alginate microparticle preparation for cell encapsulation in a microfluidic device assembled from fluoropolymer capillaries and junctions. Droplets observed at the T-junction when (b) highly viscous silicone oil (10 centistoke) and (c) low-viscosity soybean oil is used as the continuous fluid. Figure reprinted from Reference (Trivedi et al., 2010).

### III.1.2. Water insoluble or weakly soluble crosslinking agents

In the case of crosslinking agents insoluble or weakly soluble in water, mixing them with alginate in water does not lead to instant gelation since there are no available cations. In the case of crosslinking agents which are pH-sensitive, such as calcium carbonate ($CaCO_3$), an acid is used in the continuous fluid to release the cations from inert crosslinking agents. Therefore, gelation by the available cations happens after droplet generation.

In the work of Zhang et al. (Zhang et al., 2007), fine particles of $CaCO_3$ (0.1 wt%) were dispersed in an aqueous solution of Na-alginate (2 wt%). Soybean oil with a surfactant (Span 80, 3 wt%) and containing acetic acid (5 wt%) was used as the continuous fluid (Figure 13 (a)). Droplets of Na-alginate/$CaCO_3$ were generated by flow-focusing in soybean oil/acetic acid (Figure 13 (b)). Droplets pH decreased because of the acetic acid in the oil. As a result, calcium cations were released from calcium carbonate, causing internal gelation of the alginate. Finally, Ca-alginate hydrogel microparticles were collected in oil (Figure 13 (c)). However, when collected on a substrate they had a "pancake" shape and were soluble in aqueous solution owing to insufficient gelation. Moreover, no improvement was observed from increasing the concentration of acetic acid or that of $CaCO_3$. A higher concentration of $CaCO_3$ particles would give rise to their aggregation in the channel. (Zhang et al., 2007)

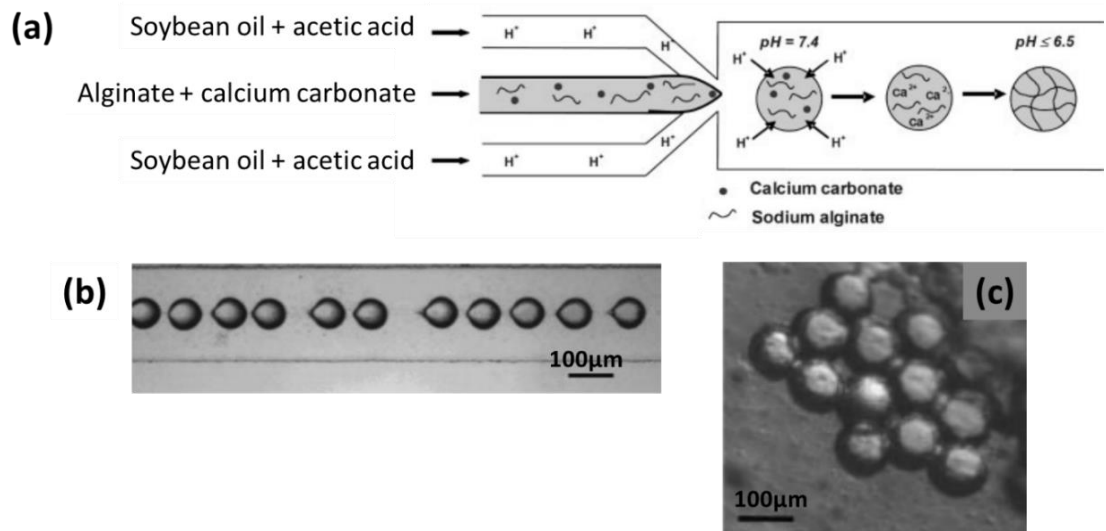

Figure 13. (a) Schematic diagram of the preparation of Ca-alginate hydrogel microparticles by using calcium carbonate to perform internal gelation of alginate in a PDMS-based microfluidic device. Micrograph of (b) droplets generated in the channel and (c) Ca-alginate hydrogel microparticles collected in oil. Figure reprinted with permission from Reference (Zhang et al., 2007)

The same principle was also applied by Akbari and Pirbodaghi to prepare cell-encapsulating microparticles (Figure 14). (Akbari and Pirbodaghi, 2014) At a first T-junction, a fluid of Na-alginate (1.5 wt%) containing cells flowed into the middle channel (Figure 14 (a)), while the Na-alginate fluid (1.5 wt%) containing $CaCO_3$ nanoparticles (35 mM) was introduced by two side channels (Figure 14 (b)). This geometry was used to create a coaxial stream while avoiding direct mechanical contact between cells and the potentially damaging $CaCO_3$ particles. At a second T-junction, fluorocarbon oil with surfactant (fluorinated surfactant, 1 wt%) was injected. Droplets of Na-alginate/cells/$CaCO_3$ were then generated by flow-focusing. After droplet collection, acetic acid (0.1 vol%) dissolved in oil was added to release calcium cations within droplets, causing gelation of alginate. Droplets were thus transformed into Ca-alginate hydrogel microparticles, some with cells encapsulated (Figure 14 (c)).

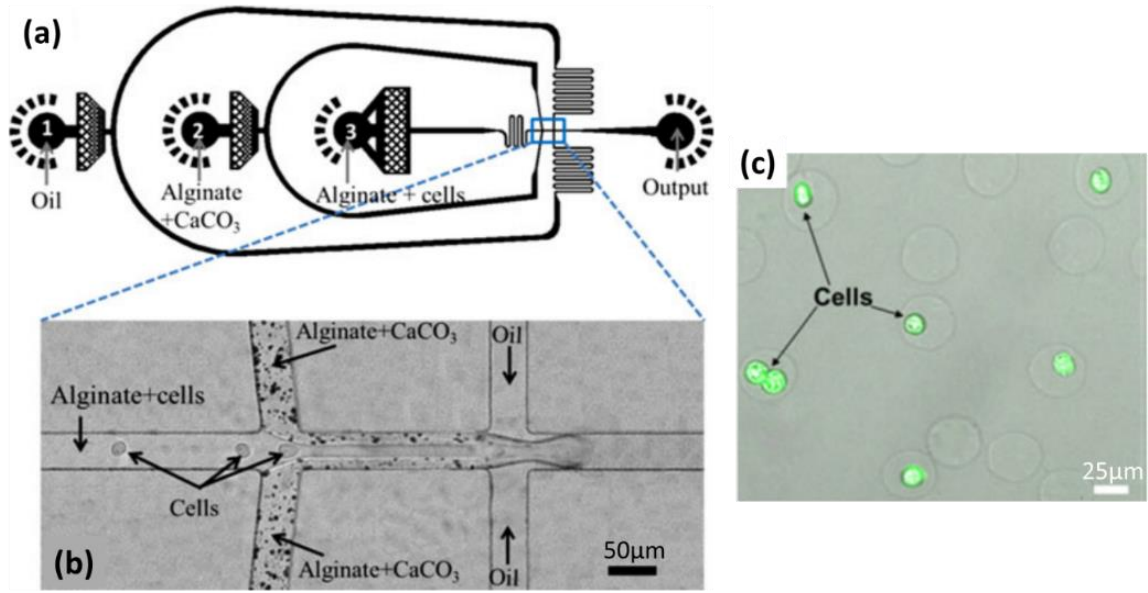

Figure 14. (a) Schematic diagram of a PDMS-based microfluidic device for the generation of droplets. (b) Micrograph of the two T-junctions in the microfluidic device. (c) Confocal microscopic image of Ca-alginate hydrogel microparticles, some with cells encapsulated (Green fluorescence represents live cells stained by calcein AM). Figure reprinted with permission from Reference (Akbari and Pirbodaghi, 2014)

In order to obtain a homogeneous internal structure of hydrogel microparticles, Utech et al. used a slightly water-soluble calcium-ethylenediaminetetraacetic acid (Ca-EDTA) complex as the crosslinking agent (Utech et al., 2015). An aqueous solution of Na-alginate (2 wt%) mixed with Ca-EDTA ($50 \times 10^{-3}$ M) was first prepared. This homogeneous mixture was used as the dispersed fluid for the microfluidic system. The continuous fluid was a fluorinated carbon oil with a biocompatible surfactant (1 wt%) containing acetic acid (0.05 vol%). Droplets of Na-alginate/Ca-EDTA were generated in oil/acetic acid by flow-focusing (Figure 15 (a)). Due to the use of acetic acid, calcium cations were released from Ca-EDTA in each droplet (Figure 15 (b)), causing internal gelation of the alginate. The Ca-alginate hydrogel microparticles formed (Figure 15 (c)) had a homogeneous internal structure and were stable in an aqueous medium without dissolution. It should be noted that, the solubility of Ca-EDTA in water being low (0.26 M at 20°C), the concentration of Ca-EDTA in the Na-alginate solution was limited in order to keep the solution homogeneous.

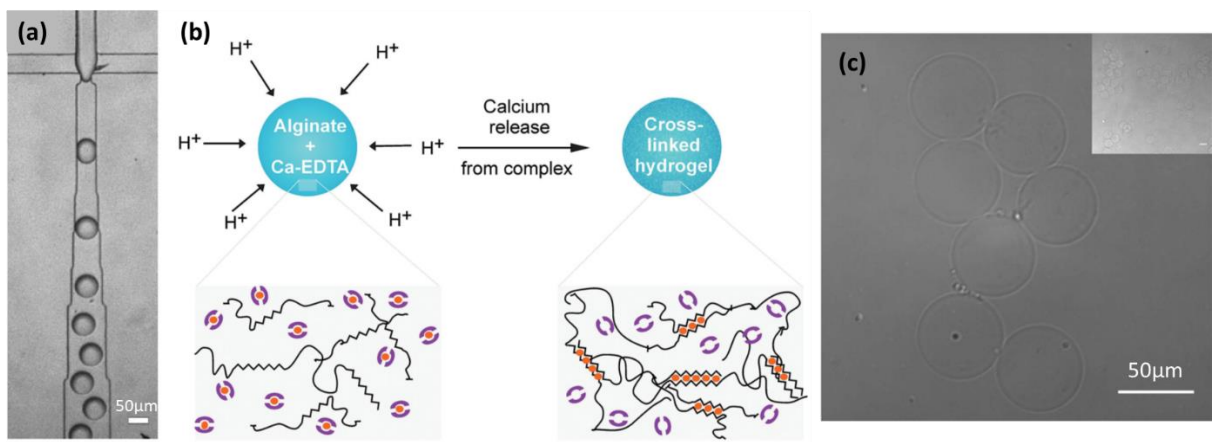

Figure 15. (a) Micrograph of the T-junction in a microfluidic device, where droplets of Na-alginate/Ca-EDTA were generated in oil/acetic acid. (b) Schematic illustration of the crosslinking process in each droplet. (c) Micrograph of Ca-alginate hydrogel microparticles in an aqueous medium. Figure reprinted with permission from Reference (Utech et al., 2015)

In conclusion, internal gelation of alginate can be realized by using crosslinking agents that are soluble or insoluble/slightly soluble in water. When water-soluble crosslinking agents are used, the instant gelation can disturb droplet generation. The problem can be solved by using partially miscible fluids with limited mixing prior to droplet generation, or by using extremely diluted solutions and surfactant.

If water-insoluble/slightly soluble crosslinking agents are used, they are mixed with alginate before droplet generation. For pH-sensitive cross-linking agents, acid is then needed to release cations, after which internal gelation takes place. A homogeneous microparticle internal structure can be achieved by choosing appropriate crosslinking agents. However, because of low solubility in water, it is important to limit the concentration of crosslinking agents to avoid precipitates in the channel.

## III.2. External gelation

In external gelation, crosslinking agents come from outside the alginate droplets and are diffused into the alginate droplets or the microparticles formed, inducing crosslinking. Unlike internal gelation, in which crosslinking agents are always introduced "on-chip" (in the microfluidic device), in external gelation, crosslinking agents can be introduced both "on-chip" and/or "off-chip" (outside the microfluidic device).

### III.2.1. On-chip introduction of crosslinking agents

Crosslinking agents can be introduced on-chip, contained in the continuous fluid, as described in Zhang et al. (Zhang et al., 2007) Calcium acetate ($Ca(CH_3COO)_2$, 2 wt%) was dissolved in soybean oil, the continuous fluid. In the microfluidic device detailed previously (III.1.2.), Na-alginate (2 wt%) droplets were generated by flow-focusing (Figure 16 (a)) in oil/calcium acetate, with surfactant (Span 80, 3 wt%). Calcium acetate diffused and dissolved in Na-alginate droplets along the channel (Figure 16 (b)), causing external gelation on-chip. Finally, Ca-alginate hydrogel microparticles were collected in oil (Figure 16 (c)). They showed better stability in an aqueous medium and had a higher Young's modulus compared with those produced by internal gelation (III.1.2.). Consequently, stronger gelation was achieved by external gelation.

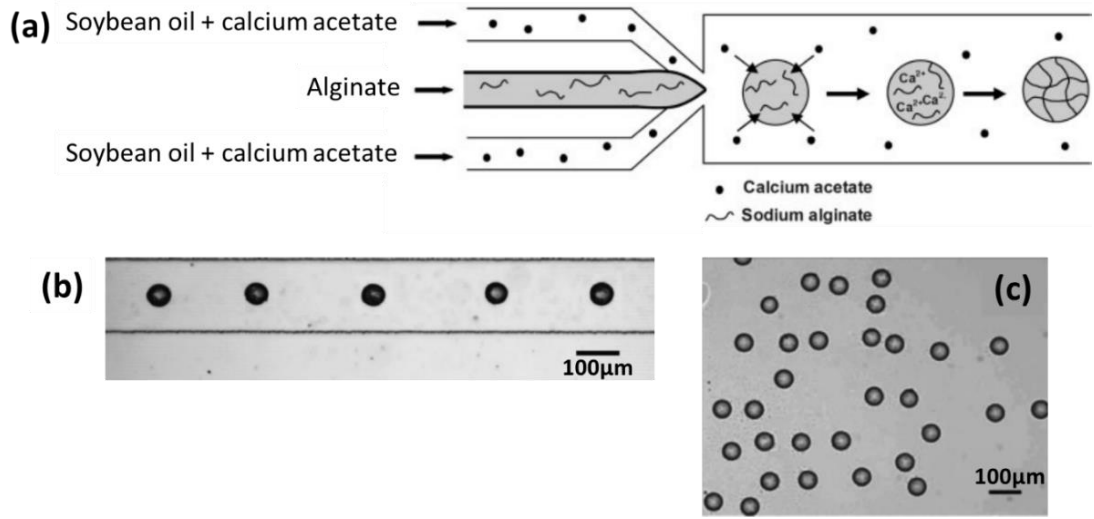

Figure 16. (a) Schematic diagram of the preparation of Ca-alginate hydrogel microparticles via on-chip external gelation in a PDMS-based microfluidic device. Micrographs of Ca-alginate hydrogel microparticles (b) in the downstream channel and (c) in the collecting container in soybean oil. Figure reprinted with permission from Reference (Zhang et al., 2007)

Smaller microparticles can be obtained without necessarily reducing channel size, by using partially miscible fluids. Sugaya et al. used methyl acetate as the continuous fluid. (Sugaya et al., 2011) Na-alginate (0.025-0.15 wt%) droplets were generated in methyl acetate (no mention of surfactant usage) by flow-focusing (Figure 17 (a)). In the following channel, because of the solubility of water in methyl acetate (8 wt%), water dissolved gradually from the droplets into methyl acetate. Thus, the droplets shrank and became more concentrated downstream. Calcium chloride solution (1M) was then injected by side channels and flowed with the droplets by co-flow. Calcium cations diffused into the droplets, inducing on-chip external gelation of alginate. Finally, Ca-alginate hydrogel microparticles with a diameter of less than 20μm were obtained (Figure 17 (b)).

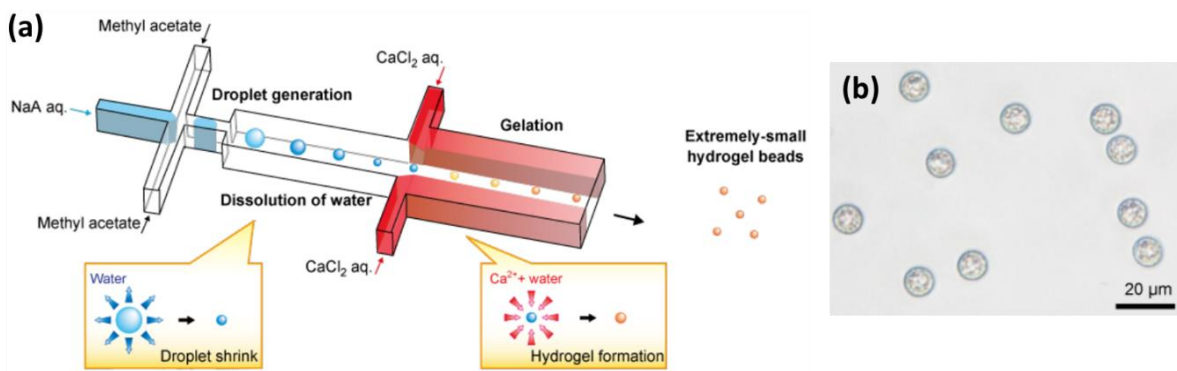

Figure 17. (a) Schematic diagram of a PDMS-based microfluidic device for the preparation of Ca-alginate hydrogel microparticles using methyl acetate as the continuous fluid, in which water is partially soluble. (b) Micrograph of Ca-alginate hydrogel microparticles collected in water. Figure reprinted from Reference (Sugaya et al., 2011)

### III.2.2. Off-chip introduction of crosslinking agents

The crosslinking agent can be introduced off-chip, as done by Hu et al. (Hu et al., 2012) to study the influence of external gelation conditions on the shape of microparticles. Na-alginate (1.5 wt%) droplets were first generated in n-decanol with surfactant (Span 80, 5 wt%), using concentric glass capillaries (Figure 18 (A)). For off-chip external gelation, droplets were collected in a two-phase gelation bath: the upper phase of n-decanol with surfactant (Span 80, 5 wt%) containing calcium chloride (15 wt%) allowed for pre-gelation of alginate; the bottom phase, an aqueous solution of barium acetate (15 wt%), strengthened the gelation. Glycerol (0-70 wt%) was added to the bottom phase to regulate viscosity. Ca-alginate hydrogel particles of different shapes (Figure 18 (B)) were obtained by varying gelation conditions such as the interfacial energy $\gamma_{CD}$, the concentration and type of surfactant, the height h between the end of the capillary and the surface of the gelation bath, and the viscosity of the bottom phase in the gelation bath. The shape of microparticles was shown to depend on forces applied to the surface of droplets when they passed through the interface in the gelation bath. The force from $\gamma_{CD}$ maintains the spherical form of droplets, while the viscous force causes deformation. The final shape resulted from the overall effect of these two forces. (Hu et al., 2012)

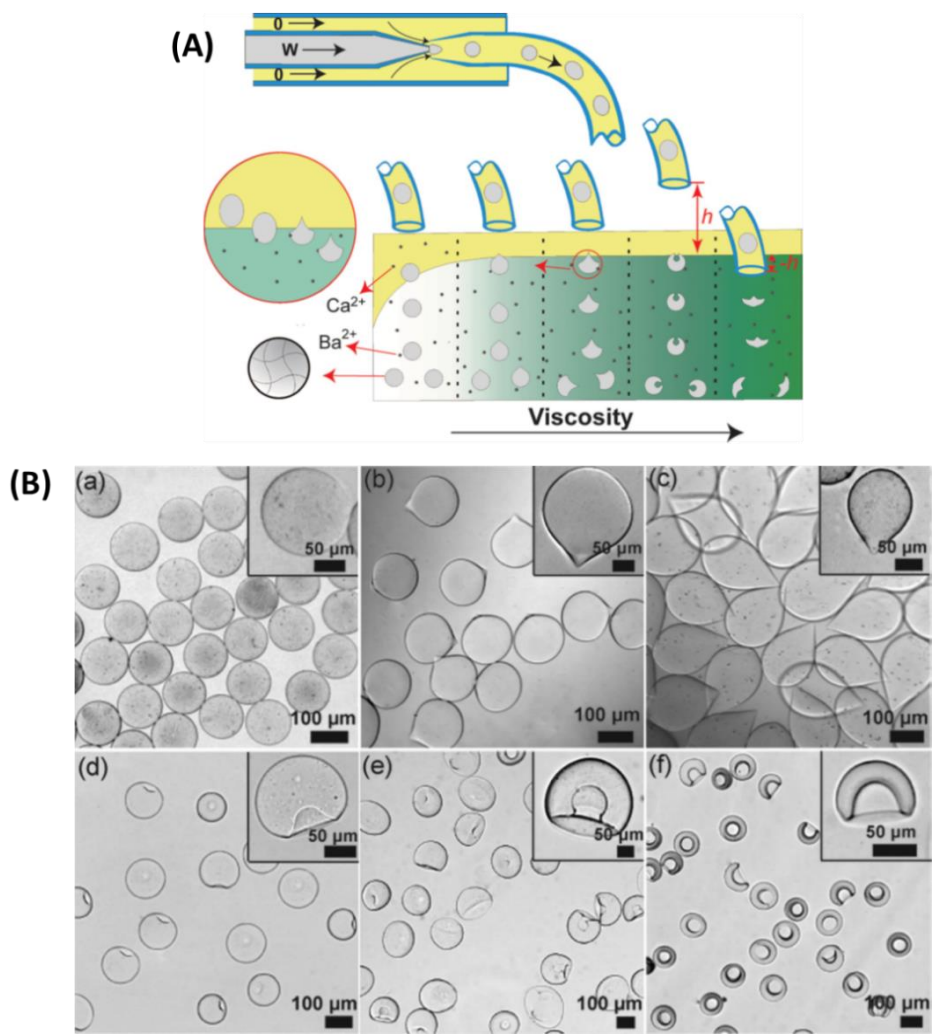

Figure 18. (A) Schematic diagram of the preparation of Ca-alginate hydrogel microparticles using a microfluidic device constructed with glass capillaries, and off-chip gelation in a two-phase gelation bath. (B) Micrographs of Ca-alginate hydrogel microparticles of different shapes prepared under different experimental conditions. Figure reprinted with permission from Reference (Hu et al., 2012)

In our group, we also collected Na-alginate droplets in an aqueous solution containing calcium chloride (Figure 19 (a)) for off-chip introduction of the crosslinking agent, but without pre-gelation. Na-alginate (0.06 wt%) droplets were first generated in dimethyl carbonate (DMC) in a T-junction, without using surfactant. The channel outlet (point B in Figure 19 (a)) was immersed in an aqueous solution of calcium chloride (0.1-1 wt%). An interface was created at the channel outlet (Figure 19 (b)) because of the non-total miscibility between DMC and water. Na-alginate droplets passed through the interface and entered the calcium chloride solution, leading to off-chip external gelation. After gelation, Ca-alginate hydrogel microparticles were droplet-shaped (Figure 19 (c)) and tadpole-shaped (Figure 19 (d)), as in Figure 18 (B, b-c). The shape of the microparticles varied with the flow rates, the concentration of Na-alginate and that of calcium chloride. It is likely that the deformation mechanism involved the forces applied to droplets at the interface, as explained by Hu et al. (Hu et al., 2012)

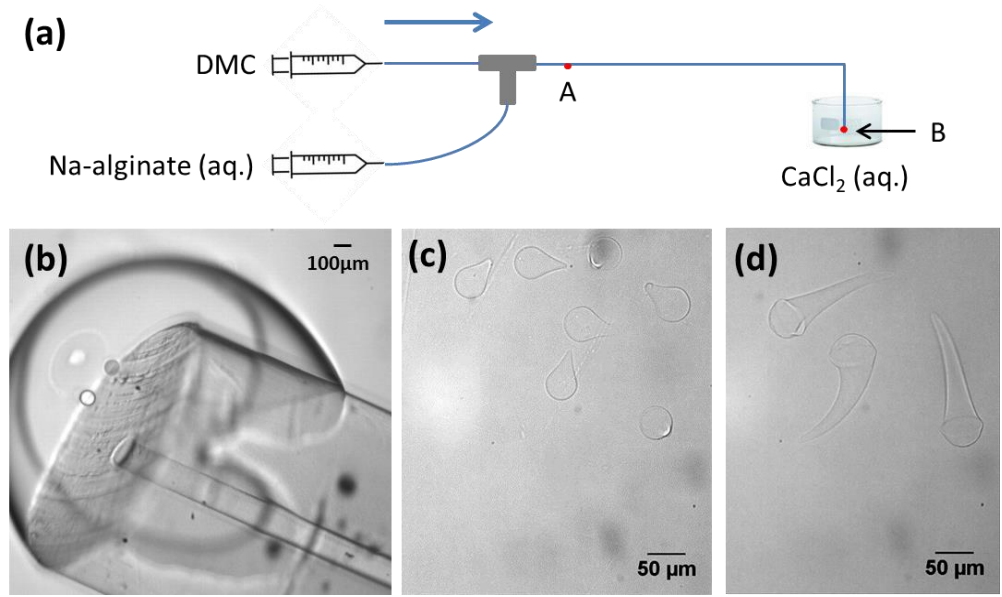

Figure 19. (a) Schematic diagram of the preparation of Ca-alginate hydrogel microparticles by off-chip external gelation without pre-gelation. Droplets were generated using a microfluidic device assembled from fluoropolymer capillaries and a T-junction. Micrographs of (b) the channel outlet immersed in an aqueous solution of calcium chloride; Ca-alginate hydrogel microparticles prepared by collecting droplets in an aqueous solution of calcium chloride at concentrations of (c) 1 wt% and (d) 0.1 wt%.

Off-chip external gelation can also be performed as a separate step after extraction of Na-alginate microparticles from the collecting bath. In Zhang et al. (Zhang et al., 2020), Na-alginate (0.006-1 wt%) droplets were generated in dimethyl carbonate (DMC) without surfactant by a T-junction (Figure 20 (a)). Because water is slightly soluble in DMC, 3 wt%, water diffused gradually from droplets into DMC, causing the droplets to shrink as they passed through the channel. Furthermore, since alginate dissolution in the continuous fluid can be ignored (Rondeau and Cooper-White, 2008), with the loss of water, the alginate concentration in droplets increased. Collected in DMC, droplets continued to shrink and were finally transformed into spherical condensed Na-alginate microparticles. After evaporation of DMC in air, an aqueous solution of calcium chloride (0.5-10 wt%) was added to the dried Na-alginate microparticles, inducing off-chip external gelation. Observations showed that this process was accompanied by the swelling of the microparticles without deformation (Figure 20 (b-c)). In the end, spherical Ca-alginate hydrogel microparticles were obtained. They were insoluble in water, indicating efficient gelation. Moreover, the concentration of calcium chloride had no significant effect on the size of the Ca-alginate microparticles.

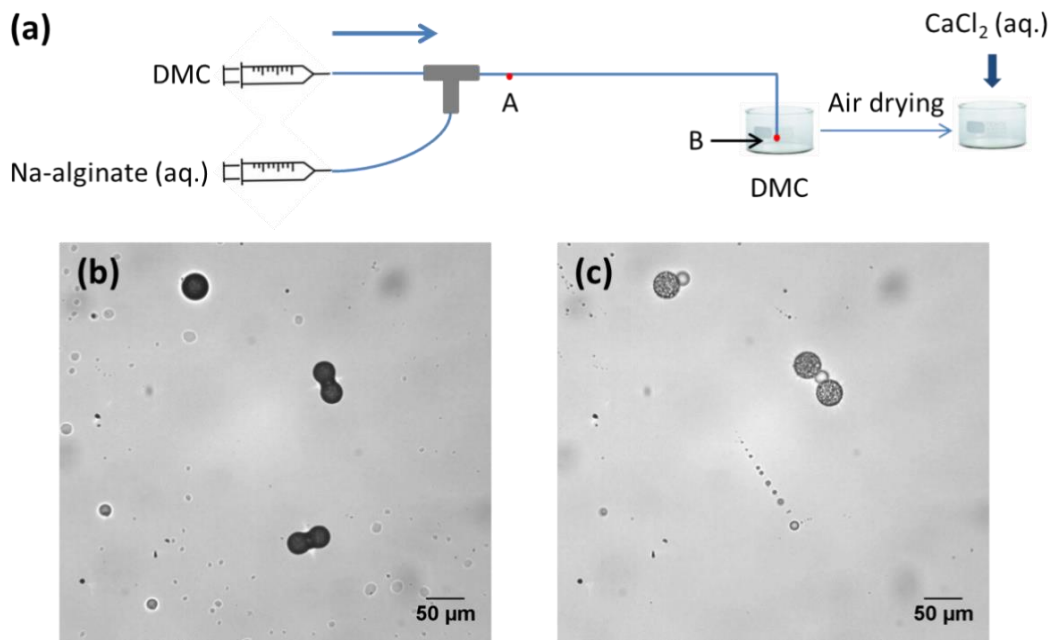

Figure 20. (a) Schematic diagram of a two-step preparation of Ca-alginate hydrogel microparticles using a microfluidic device constructed from fluoropolymer capillaries and junctions. Micrographs of (b) dried Na-alginate microparticles in air and (c) corresponding Ca-alginate microparticles after gelation in calcium chloride solution. Figure reprinted and adapted with permission from Reference (Zhang et al., 2020)

In conclusion, external gelation of alginate can be performed both on-chip and off-chip. For on-chip external gelation, crosslinking agents can be added in the continuous fluid, i.e. the oil. However, as for ionic crosslinking (Chapter I.2), limited concentrations of crosslinking agents can be used, since most of them are slightly soluble in oil. On the other hand, if partially miscible fluids are used, after droplet shrinkage in oil, aqueous solutions of crosslinking agents can be injected downstream and flow coaxially with droplets.

For off-chip external gelation, since crosslinking agents can be dissolved in an aqueous solution, their concentrations can vary over a much larger range. Ca-alginate microparticles can be deformed owing to the non-total miscibility between the oil phase and the aqueous solution. However, it remains possible to produce spherical microparticles by controlling experimental conditions and methods. Thus, off-chip external gelation can be used to produce shape-controlled microparticles.

## IV. Properties of alginate hydrogel microparticles

After preparation, alginate hydrogel microparticles should be characterized to obtain better knowledge of their properties, which will determine their further applications. This section discusses characterization approaches and factors influencing particle properties.

## IV.1. Size

Size is one of the most important properties of alginate hydrogel microparticles. For example, in drug delivery, microparticle size and size distribution affect drug release kinetics. (Uyen et al., 2020) Size can be measured by optical or light-scattering (sub-micrometer range) microscopy (Joye and Mcclements, 2014), or using microgrippers. (Zhang, 2020)

Droplet-based microfluidics allows monodisperse microparticles to be produced with accurate control of size and size distribution. Table 1 shows the average size attained under droplet-based microfluidics using different gelation methods. Figure 21 shows an example of the narrow size distribution of Na-alginate and Ca-alginate microparticles produced by droplet-based microfluidics (Zhang et al., 2020), indicating the monodispersity of the particle size. This is an advantage compared to conventional emulsification, which yields a broader size distribution. (Xu et al., 2009)

Table 1. Average size of alginate microparticles prepared using droplet-based microfluidics with different gelation methods.

| Average size of microparticles | Gelation method | Reference |
|---|---|---|
| 1-50 µm<br>10-300 nm | Internal gelation with water-soluble crosslinking agents mixed with Na-alginate before droplet generation | Rondeau and Cooper-White 2008* |
| 50-300 µm | Internal gelation with water-soluble crosslinking agents mixed with Na-alginate after droplet generation | Xu et al. 2008 |
| 20-50 µm | | Liu et al. 2006 |
| 22-42 µm | | Trivedi et al. 2009 |
| 60-100 µm | Internal gelation with water-insoluble crosslinking agents mixed with Na-alginate after droplet generation | Zhang et al. 2007 |
| 26 µm | | Akbari and Pirbodaghi 2014 |
| 10-50 µm | Internal gelation with slightly water-soluble crosslinking agents mixed with Na-alginate after droplet generation | Utech et al. 2015 |
| 50-70 µm | External gelation with on-chip introduction of crosslinking agents | Zhang et al. 2007 |
| 6-10 µm | | Sugaya et al. 2011* |
| 100-200 µm | External gelation with off-chip introduction of crosslinking agents | Hu et al. 2012 |

| 7-40 μm | Zhang et al. 2020* |

\* Partially miscible fluids were used.

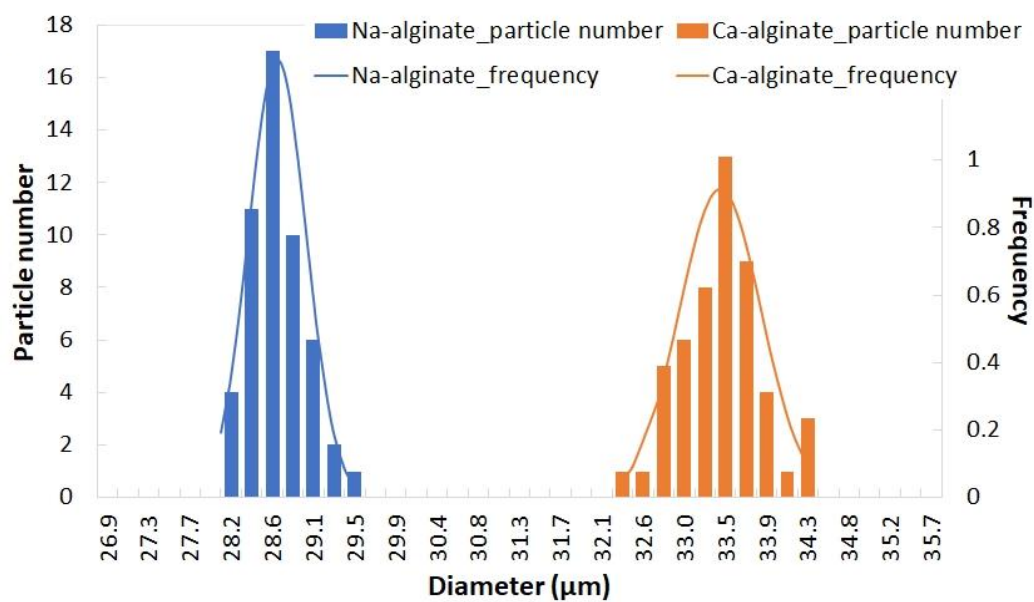

Figure 21. Size distribution of microparticles of Na-alginate (blue) and Ca-alginate (orange) produced using droplet-based microfluidics. The curves show Gaussian fitting. Figure reprinted and adapted with permission from Reference (Zhang et al., 2020)

With droplet-based microfluidics, the size of alginate hydrogel microparticles is influenced by several factors linked to the fluids used to generate them. When immiscible fluids are used, the size of alginate hydrogel microparticles is completely dependent on the size of the droplets first generated. Droplet size is influenced by channel size, and smaller droplets can be generated by using narrower channels. Other important factors are flow rates, alginate concentration (Rondeau and Cooper-White, 2008; Zhang et al., 2020), and fluid viscosities. (Seemann et al., 2011; Teh et al., 2008) However, reducing channel size increases hydraulic resistance, as well as the pressure required to generate droplets. Moreover, it should be noted that in most cases, the Na-alginate solution used is relatively viscous. Therefore, high pressure may cause leakage or even destruction of the microfluidic device. (Akbari and Pirbodaghi, 2014; Utech et al., 2015) Thus, even when channel diameter is decreased and/or the flow rate of the continuous fluid is increased, producing droplets of a diameter below 10 μm remains challenging.

Droplets of this size can be obtained without applying high pressure (Table 1), by using partially miscible fluids with low solubility in each other. (Rondeau and Cooper-White, 2008; Sugaya et al., 2011; Zhang et al., 2020) The dispersed fluid is an aqueous solution containing Na-alginate; the continuous fluid is an organic solvent that is partially miscible with water and in which water has low solubility. The partial miscibility between the continuous and the dispersed fluids should be slight enough so that interfacial energy $\gamma_{CD}$ still allows the generation of droplets. The low solubility of water in the continuous fluid allows water diffusion from droplets into it, causing the droplets to shrink. As a result, the initially-obtained diluted large droplets are transformed into concentrated small droplets or microparticles. Thus, their size is no longer dependent on the size of droplets initially generated, but varies with the interaction between water and the continuous fluid.

### IV.2. Shape

The shape of microparticles is another important property. A specific shape is sometimes needed; for example, red blood cell-mimicking microparticles are often required in a biconcave shape. (Merkel et al., 2011) In drug delivery, the shape of microparticles has an impact on the drug-release profile. (Freiberg and Zhu, 2004) The overall shape of microparticles can be observed by using optical microscopy. Confocal microscopy of fluorescent samples can be used to form a spatial 3D image. (Joye and Mcclements, 2014) Better resolution can be obtained by using atomic force microscopy (AFM) or scanning electron microscopy (SEM). (Zhang et al., 2020)

With droplet-based microfluidics, the spherical droplets initially generated can be transformed into spherical alginate hydrogel microparticles after gelation. Non-spherical microparticles can also be obtained. For example, as presented previously, Liu et al. first generated droplets of Na-alginate and calcium chloride separately. (Liu et al., 2006) Then the droplets were fused in a specifically designed microfluidic device, leading to gelation. By varying the channel geometry and controlling the flow rates of fluids, Ca-alginate microparticles of different shapes were obtained (Figure 11 (f)). A different method was presented by Hu et al. (Hu et al., 2012) Na-alginate droplets were first generated and then collected in a two-phase gelation bath. Spherical droplets were deformed via interfacial energy derived from surfactant and viscous force. Thus, different shapes were produced (Figure 18 (B)) by controlling the surfactant used and the viscosity of the gelation bath.

### IV.3. Concentration

After preparation, the concentration of alginate in the microparticles can be calculated approximately. For instance, Zhang et al. used partially miscible fluids. (Zhang et al., 2020) An aqueous solution of Na-alginate was prepared with a known concentration. After droplet generation, droplet shrinkage occurred during passage through the channel due to water diffusion into the continuous fluid. Droplets were hence transformed into microparticles. As the diffusion of Na-alginate into the continuous fluid is negligible (Rondeau and Cooper-White, 2008), the quantity of Na-alginate is constant. It can be calculated by multiplying the droplet volume and initial concentration. Finally, by measuring microparticle size, the concentration of Na-alginate can be calculated. The final concentration of Na-alginate varies from 20 to 100 wt%, depending on the initial concentration and diameter of the droplets generated. (Zhang et al., 2020) Furthermore, as presented in Utech et al., the homogeneity of composition of microparticles can be determined with the help of fluorescence technology. (Utech et al., 2015)

### IV.4. Stability

In most cases, surfactant is added in the continuous fluid (Trivedi et al., 2009; Utech et al., 2015; Zhang et al., 2007) to lower interfacial energy $\gamma_{CD}$. Note that for each of the above studies, use or non-use of surfactant is mentioned when indicated by the authors. Surfactant facilitates the creation of a new interface, and thus the formation of droplets. It also stabilizes the formed droplets by preventing their coalescence. (Seemann et al., 2011) Before the application of microparticles, the surfactant should be dissolved (Akbari and Pirbodaghi, 2014), except for biocompatible surfactant (Utech et al., 2015), although protocols for removing surfactant are rarely reported in the literature. To remove the oil used during the preparation, microparticles should be washed several times with an aqueous solution, followed by centrifugation. (Zhang et al., 2007)

However, despite its advantages, the use of surfactant may be undesirable. Surfactant has been shown to impact the surface properties of microparticles, such as morphology (Sundberg et al., 1990) and surface hydrophobicity. (Kidane et al., 2002) Additionally, if rinsing is insufficient, the traces of surfactant on microparticles can damage the devices during application. In this situation, microparticles should be prepared without surfactant, which is possible with droplet-based microfluidics. In the microchannel, the coalescence of droplets can be avoided by enlarging the distance between droplets, which can be achieved by regulating flow rates. Furthermore, gelation, either on-chip or off-chip, solidifies droplets and thus helps to avoid coalescence as well. Another strategy consists of using partially miscible fluids. This means that the droplets shrink and become more and more condensed during passage through the channel. At the outlet, either gelation (Rondeau and Cooper-White, 2008) or a final shrinkage (Zhang et al., 2020) can help avoid coalescence.

Moreover, for their stability, alginate hydrogel microparticles should be insoluble in water. This can be achieved by adopting proper gelation methods using a sufficient quantity of crosslinking agents for effective gelation.

## IV.5. Mechanical properties

Mechanical properties of alginate hydrogel microparticles are usually characterized by measuring the Young's modulus, which varies with several factors. According to the type of bond between alginate and crosslinking agents, covalent crosslinking results in a higher Young's modulus than ionic crosslinking in microparticles. (Caccavo et al., 2018) For ionic crosslinking, different cations present different affinities, i.e. different forces with alginate, thus different Young's moduli. (Mørch et al., 2006) In addition, the Young's modulus increases with the concentration of alginate. (Markert et al., 2013) To measure the Young's modulus of a microparticle, it needs to be deformed under a known force, which can be either compressive or tensile. (Guevorkian and Maître, 2017) The techniques used in the literature include micropipette aspiration (Kleinberger et al., 2013), compression (Carin et al., 2003; Wang et al., 2005) or Atomic Force Microscopy (AFM). (Zhang et al., 2020; Zhang et al., 2007)

### IV.5.1. Micropipette aspiration technique

In the micropipette aspiration technique, controlled pressure is used to pull on the sample surface. When this pressure is high enough, the sample behaves like a viscoelastic fluid flowing inside the micropipette. (Guevorkian and Maître, 2017) With a known pressure applied, the Young's modulus is calculated via equations based on corresponding models. (Kleinberger et al., 2013)

### IV.5.2. Compression technique

This technique consists in compressing a microparticle between two parallel plates (Carin et al., 2003) or between the flat end of a glass fiber and a glass surface. (Wang et al., 2005) A force transducer is connected to the equipment to measure the force applied. By varying the force, microparticle deformation can be recorded. Finally, according to the force-deformation curve and equations based on theoretical models, the Young's modulus is calculated.

However, both the micropipette aspiration technique and the compression technique are unsuitable for microparticles with high resistance to deformation. (Zhang et al., 2020) In this case, Atomic Force Microscopy can be used.

### IV.5.3. Atomic Force Microscopy

For Atomic Force Microscopy (AFM), the indentation depth (order of 100 nm) is generally about 100 times less than the diameter of the microparticle (order of 10 µm). Hence the Young's modulus represents the local mechanical property on the surface, depending on the measuring point, as in Zhang et al. (Zhang et al., 2020) As the surface of their microparticles was smooth, variation in the local mechanical property was explained by the porous inner structure observed by Scanning Electron Microscopy (SEM) (Figure 22).

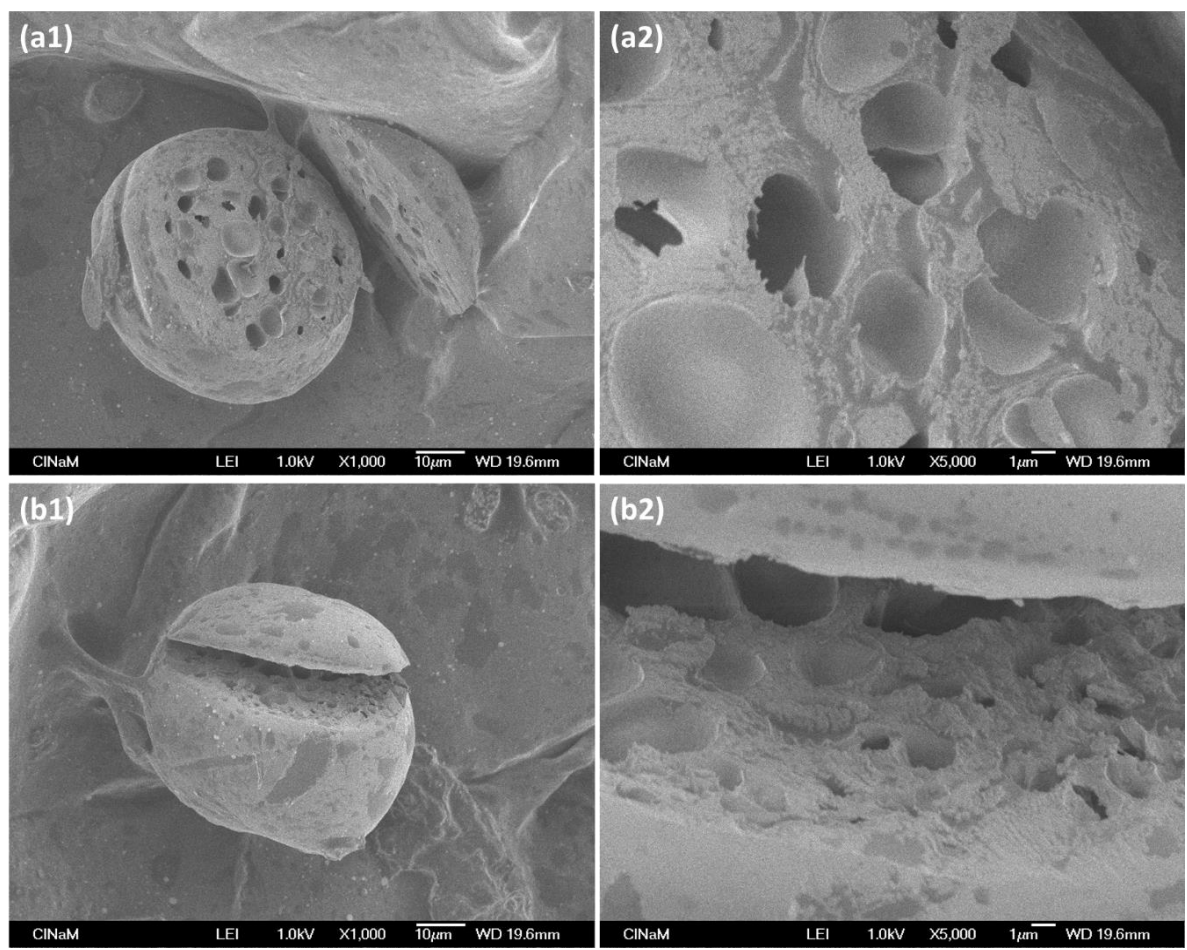

Figure 22. SEM photographs of 2 Na-alginate microparticles (a) and (b), magnified 1000x (a1 and b1) and 5000x (a2 and b2). Na-alginate microparticles were prepared following the method mentioned in the publication (Zhang et al. 2020). Figure reprinted and adapted with permission from Reference (Zhang et al., 2020)

# V. Conclusion

This review focuses on a microparticle-producing technique widely used for its efficacy in controlling physicochemical properties: droplet-based microfluidics.

To transform alginate microdroplets into alginate hydrogel microparticles, gelation is indispensable. It is realized by crosslinking, which requires crosslinking agents to be introduced either inside or outside the microdroplets, causing respectively internal gelation or external gelation. In external gelation, crosslinking agents can be introduced both "on-chip" and/or "off-chip" (outside the microfluidic device). The review describes the various strategies applied under this technique, and the size, shape, concentration, stability and mechanical properties of the alginate hydrogel microparticles obtained.

# IV. References

Agüero, L., Zaldivar-Silva, D., Peña, L., Dias, M.L., 2017. Alginate microparticles as oral colon drug delivery device: A review. Carbohydrate Polymers 168, 32-43.

Ahmed, E.M., 2015. Hydrogel: Preparation, characterization, and applications: A review. Journal of Advanced Research 6, 105-121.

Akbari, S., Pirbodaghi, T., 2014. Microfluidic encapsulation of cells in alginate particles via an improved internal gelation approach. Microfluidics and Nanofluidics 16, 773-777.

Baroud, C.N., Gallaire, F., Dangla, R., 2010. Dynamics of microfluidic droplets. Lab on a Chip 10, 2032-2045.

Caccavo, D., Cascone, S., Lamberti, G., Barba, A.A., 2018. Hydrogels: experimental characterization and mathematical modelling of their mechanical and diffusive behaviour. Chemical Society Reviews 47, 2357-2373.

Carin, M., Barthès-Biesel, D., Edwards-Lévy, F., Postel, C., Andrei, D.C., 2003. Compression of biocompatible liquid-filled HSA-alginate capsules: Determination of the membrane mechanical properties. Biotechnology and Bioengineering 82, 207-212.

Chan, L.W., Lee, H.Y., Heng, P.W.S., 2002. Production of alginate microspheres by internal gelation using an emulsification method. International Journal of Pharmaceutics 242, 259-262.

Christopher, G.F., Anna, S.L., 2007. Microfluidic methods for generating continuous droplet streams. Journal of Physics D: Applied Physics 40, R319-R336.

De Menech M., Garstecki P., Jousse F., Stone H. A., 2008. Transition from squeezing to dripping in a microfluidic T-shaped junction. , 595, pp 141-161 Journal of Fluid Mechanics 595, 141-161.

Draget, K.I., 2009. 29 - Alginates, in: Phillips, G.O., Williams, P.A. (Eds.), Handbook of Hydrocolloids (Second Edition). Woodhead Publishing, pp. 807-828.

Freiberg, S., Zhu, X.X., 2004. Polymer microspheres for controlled drug release. International Journal of Pharmaceutics 282, 1-18.

Fundueanu, G., Nastruzzi, C., Carpov, A., Desbrieres, J., Rinaudo, M., 1999. Physico-chemical characterization of Ca-alginate microparticles produced with different methods. Biomaterials 20, 1427-1435.

Gacesa, P., 1988. Alginates. Carbohydrate Polymers 8, 161-182.

Grant, G.T., Morris, E.R., Rees, D.A., Smith, P.J.C., Thom, D., 1973. Biological interactions between polysaccharides and divalent cations: The egg-box model. FEBS Letters 32, 195-198.

Guevorkian, K., Maître, J.L., 2017. Chapter 10 - Micropipette aspiration: A unique tool for exploring cell and tissue mechanics in vivo, in: Lecuit, T. (Ed.), Methods in Cell Biology. Academic Press, pp. 187-201.

Haghgooie, R., Toner, M., Doyle, P.S., 2010. Squishy Non-Spherical Hydrogel Microparticles. Macromolecular Rapid Communications 31, 128-134.

Hoffman, A.S., 2012. Hydrogels for biomedical applications. Advanced Drug Delivery Reviews 64, 18-23.

Hu, Y., Wang, Q., Wang, J., Zhu, J., Wang, H., Yang, Y., 2012. Shape controllable microgel particles prepared by microfluidic combining external ionic crosslinking. Biomicrofluidics 6, 026502.

Joye, I.J., Mcclements, D.J., 2014. Biopolymer-based nanoparticles and microparticles: Fabrication, characterization, and application. Current Opinion in Colloid & Interface Science 19, 417-427.

Kidane, A., Guimond, P., Rob Ju, T.-C., Sanchez, M., Gibson, J., North, A., Hogenesch, H., Bowersock, T.L., 2002. Effects of cellulose derivatives and poly(ethylene oxide)–poly(propylene oxide) tri-block copolymers (Pluronic®surfactants) on the properties of alginate based microspheres and their interactions with phagocytic cells. Journal of Controlled Release 85, 181-189.

Kleinberger, R.M., Burke, N.a.D., Dalnoki-Veress, K., Stöver, H.D.H., 2013. Systematic study of alginate-based microcapsules by micropipette aspiration and confocal fluorescence microscopy. Materials Science and Engineering: C 33, 4295-4304.

Lee, B.-B., Ravindra, P., Chan, E.-S., 2013. Size and Shape of Calcium Alginate Beads Produced by Extrusion Dripping. Chemical Engineering & Technology 36, 1627-1642.

Lee, K.Y., Mooney, D.J., 2012. Alginate: Properties and biomedical applications. Progress in Polymer Science 37, 106-126.

Lee, K.Y., Yuk, S.H., 2007. Polymeric protein delivery systems. Progress in Polymer Science 32, 669-697.

Liu, K., Ding, H.-J., Liu, J., Chen, Y., Zhao, X.-Z., 2006. Shape-Controlled Production of Biodegradable Calcium Alginate Gel Microparticles Using a Novel Microfluidic Device. Langmuir 22, 9453-9457.

Maitra, J., Shukla, V., 2014. Cross-linking in hydrogels - a review. Am J Polym Sci 4, 25-31.

Markert, C.D., Guo, X., Skardal, A., Wang, Z., Bharadwaj, S., Zhang, Y., Bonin, K., Guthold, M., 2013. Characterizing the micro-scale elastic modulus of hydrogels for use in regenerative medicine. Journal of the Mechanical Behavior of Biomedical Materials 27, 115-127.

Merkel, T.J., Jones, S.W., Herlihy, K.P., Kersey, F.R., Shields, A.R., Napier, M., Luft, J.C., Wu, H., Zamboni, W.C., Wang, A.Z., Bear, J.E., Desimone, J.M., 2011. Using mechanobiological mimicry of red blood cells to extend circulation times of hydrogel microparticles. Proceedings of the National Academy of Sciences 108, 586.

Mørch, Ý.A., Donati, I., Strand, B.L., Skjåk-Bræk, G., 2006. Effect of $Ca^{2+}$, $Ba^{2+}$, and $Sr^{2+}$ on Alginate Microbeads. Biomacromolecules 7, 1471-1480.

Qiu, C., Chen, M., Yan, H., Wu, H., 2007. Generation of Uniformly Sized Alginate Microparticles for Cell Encapsulation by Using a Soft-Lithography Approach. Advanced Materials 19, 1603-1607.

Ren, K., Zhou, J., Wu, H., 2013. Materials for Microfluidic Chip Fabrication. Accounts of Chemical Research 46, 2396-2406.

Rondeau, E., Cooper-White, J.J., 2008. Biopolymer Microparticle and Nanoparticle Formation within a Microfluidic Device. Langmuir 24, 6937-6945.

Santa-Maria, M., Scher, H., Jeoh, T., 2012. Microencapsulation of bioactives in cross-linked alginate matrices by spray drying. Journal of Microencapsulation 29, 286-295.

Seemann, R., Brinkmann, M., Pfohl, T., Herminghaus, S., 2011. Droplet based microfluidics. Reports on Progress in Physics 75, 016601.

Stephenson, R., Stuart, J., 1986. Mutual binary solubilities: water-alcohols and water-esters. Journal of Chemical and Engineering Data 31, 56-70.

Sugaya, S., Yamada, M., Seki, M., 2011. Production of extremely-small hydrogel microspheres by utilizing water-droplet dissolution in a polar solvent, in: Landers, J. (Ed.), 15th International Conference on Miniaturized Systems for Chemistry and Life Sciences. Chemical and Biological Microsystems Society ( CBMS ), Seattle, Washington, USA, pp. 18-20.

Sundberg, D.C., Casassa, A.P., Pantazopoulos, J., Muscato, M.R., Kronberg, B., Berg, J., 1990. Morphology development of polymeric microparticles in aqueous dispersions. I. Thermodynamic considerations. Journal of Applied Polymer Science 41, 1425-1442.

Teh, S.Y., Lin, R., Hung, L.H., Lee, A.P., 2008. Droplet microfluidics. Lab on a Chip 8, 198-220.

Trivedi, V., Doshi, A., Kurup, G.K., Ereifej, E., Vandevord, P.J., Basu, A.S., 2010. A modular approach for the generation, storage, mixing, and detection of droplet libraries for high throughput screening. Lab on a Chip 10, 2433-2442.

Trivedi, V., Ereifej, E.S., Doshi, A., Sehgal, P., Vandevord, P.J., Basu, A.S., 2009. Microfluidic encapsulation of cells in alginate capsules for high throughput screening, Proceedings of the 31st Annual International Conference of the IEEE Engineering in Medicine and Biology Society: Engineering the Future of Biomedicine, EMBC 2009, pp. 7037-7040.

Utech, S., Prodanovic, R., Mao, A.S., Ostafe, R., Mooney, D.J., Weitz, D.A., 2015. Microfluidic Generation of Monodisperse, Structurally Homogeneous Alginate Microgels for Cell Encapsulation and 3D Cell Culture. Advanced Healthcare Materials 4, 1628-1633.

Uyen, N.T.T., Hamid, Z.a.A., Tram, N.X.T., Ahmad, N., 2020. Fabrication of alginate microspheres for drug delivery: A review. International Journal of Biological Macromolecules 153, 1035-1046.

Velings, N.M., Mestdagh, M.M., 1995. Physico-chemical properties of alginate gel beads. Polymer Gels and Networks 3, 311-330.

Wang, C.X., Cowen, C., Zhang, Z., Thomas, C.R., 2005. High-speed compression of single alginate microspheres. Chemical Engineering Science 60, 6649-6657.

Xu, J.H., Li, S.W., Tan, J., Luo, G.S., 2008. Controllable Preparation of Monodispersed Calcium Alginate Microbeads in a Novel Microfluidic System. Chemical Engineering & Technology 31, 1223-1226.

Xu, Q., Hashimoto, M., Dang, T.T., Hoare, T., Kohane, D.S., Whitesides, G.M., Langer, R., Anderson, D.G., 2009. Preparation of Monodisperse Biodegradable Polymer Microparticles Using a Microfluidic Flow-Focusing Device for Controlled Drug Delivery. Small 5, 1575-1581.

Zagnoni, M., Anderson, J., Cooper, J.M., 2010. Hysteresis in Multiphase Microfluidics at a T-Junction. Langmuir 26, 9416-9422.

Zhang, C., 2020. Development of a microfluidic method for the preparation of mimetic microparticles of red blood cells with controllable size and mechanical properties. Université d'Aix-Marseille.

Zhang, C., Grossier, R., Lacaria, L., Rico, F., Candoni, N., Veesler, S., 2020. A microfluidic method generating monodispersed microparticles with controllable sizes and mechanical properties. Chemical Engineering Science 211, 115322.

Zhang, H., Tumarkin, E., Peerani, R., Nie, Z., Sullan, R.M.A., Walker, G.C., Kumacheva, E., 2006. Microfluidic Production of Biopolymer Microcapsules with Controlled Morphology. Journal of the American Chemical Society 128, 12205-12210.

Zhang, H., Tumarkin, E., Sullan, R.M.A., Walker, G.C., Kumacheva, E., 2007. Exploring Microfluidic Routes to Microgels of Biological Polymers. Macromolecular Rapid Communications 28, 527-538.

Zhang, S., Guivier-Curien, C., Veesler, S., Candoni, N., 2015. Prediction of sizes and frequencies of nanoliter-sized droplets in cylindrical T-junction microfluidics. Chemical Engineering Science 138, 128-139.

Zhu, P., Wang, L., 2017. Passive and active droplet generation with microfluidics: a review. Lab on a Chip 17, 34-75.